\documentclass{article}
\usepackage{xcolor}

\usepackage{type1cm}        
\usepackage{makeidx}         
\usepackage{graphicx}        
\usepackage{multicol}        
\usepackage[bottom]{footmisc}

\usepackage{newtxtext}       %
\usepackage{newtxmath}       

\usepackage{subfig}
\usepackage{enumerate}
\usepackage{calrsfs}
\usepackage{graphicx}

\makeindex             

\newcommand{\bx}{\mathbf{x}}
\newcommand{\f}{\mathbf{f}}
\newcommand{\df}{\mathbf{df}}

\begin{document}

\title{Topological Feature Search in Time-Varying Multifield Data}

\author{Tripti Agarwal \thanks{ International Institute of Information Technology, Bangalore, India. {\tt tripti.agarwal@iiitb.org}
                      }
\and
Amit Chattopadhyay \thanks{International Institute of Information Technology, Bangalore, India. {\tt
            a.chattopadhyay@iiitb.ac.in}}
\and
Vijay Natarajan\thanks{Indian Institute of Science, Bangalore, India. {\tt
            vijayn@iisc.ac.in}}
}
%
%
\date{}
\maketitle

\begin{abstract}
A wide range of data that appear in scientific experiments
  and simulations are multivariate or multifield in nature, consisting
  of multiple scalar fields. Topological feature search of such data
  aims to reveal important properties useful to the domain
  scientists. It has been  shown in recent works that a single scalar
  field is insufficient to capture many important topological features
  in the data, instead one needs to consider topological relationships
  between multiple scalar fields. In the current paper, we propose a
  novel method of finding similarity between two multifield data by
  comparing their respective fiber component distributions. Given a
  time-varying multifield data, the method computes a metric plot for
  each pair of histograms at consecutive time stamps to understand the
  topological changes in the data over time. We validate the method
  using real and synthetic data. The effectiveness of the proposed
  method is shown by its ability to capture important topological
  features that are not always possible to detect using the individual
  component scalar fields.
\end{abstract}




\section{Introduction}
\label{sec:intro}
Scientists understand different physical phenomena by studying the interrelationships between features in different fields. It has been observed and shown that such multifield or multivariate data can reveal many important phenomena about an experiment that are impossible to study using a single scalar field data~\cite{2012-Duke-VisWeek, 2015-Carr-Fiber}. Development of tools and techniques for extracting and visualizing features in multifield data is an important topic of research interest~\cite{Dagstuhl-2014}. Topology-based methods have been shown to be extremely effective in this context. During the previous two decades, topological analysis of shapes and data was mostly driven by scalar topology, using contour tree, Reeb graph, Morse-Smale complex and their variants~\cite{Biasotti2008DescribingSB}. Such techniques have also been extended for time-varying scalar field data by defining different topology-aware similarity measures between two scalar fields~\cite{saikia2014extended, narayanan2015distance, bauer2014measuring}.

Generalization of the techniques to time-varying multifield data is challenging and requires further development in both theory and computational methods. More recently, new tools have been proposed for understanding and visualizing multifield data --  Reeb Space~\cite{2008-edels-reebspace}, Jacobi set~\cite{2004-edels-jacobi,2007-Bremer,2004-edels-localglobal}, Joint Contour Net~\cite{JCN_paper,2012-Duke-VisWeek} and Pareto analysis~\cite{2013-Huettenberger-pareto}. Extending these methods to time-varying multifield data requires the development of techniques for comparative analysis and visualization. For example, developing a comparative measure between two Reeb spaces is a challenging open problem. In this paper, we consider a simpler feature descriptor of a multifield, namely its fiber-component distribution or histogram.  Using this, we make a first step forward towards a topology-aware distance measure between two multifields in terms of  the distance between their fiber-component distributions. Our contribution in the current paper is as follows:

\begin{itemize}
\item We introduce simple topology-aware distance measures between two multifields based on their fiber-component distributions or histograms in the range space. We prove the metric properties of the proposed distance measures.
    
\item We show that the proposed measures capture significant or interesting events in time-varying phenomena, not possible using a study of individual fields. We validate the method by experimenting on a time-varying synthetic data where topological features are known in advance. 

\item We show effectiveness of our method by experimenting on previously studied nuclear-scission data~\cite{2012-Duke-VisWeek} and re-explain how scission events are captured. We also apply our method in capturing important feature in the orbital data of Pt-CO interaction.
\end{itemize}

\noindent
Section 2  discusses related works on scalar and multifield data analysis. Section 3 describes different data structures or representations used for understanding and visualizing multifield data. Section 4 introduces our proposed topology-aware distance measures and describes important properties of the measure. Section 5 discusses the implementation details and Section 6 and Section 7 describe various results of experiments on synthetic and real data. The experiments are conducted on nuclear scission, fission, and molecular orbital density data of Pt-CO interaction.
Finally, Section 8 presents conclusions and lists some limitations of the method.
\section{Related Work}
\label{sec:related}
Feature extraction in time-varying data is a well studied topic and several approaches have been proposed. We describe a few relevant approaches here. 

Various similarity measures between scalar fields have been studied to analyze repeating patterns and similar arrangements in the data. Hilaga et al. studied topological shape matching using a multiresolution Reeb Graph (MRG)~\cite{hilaga2001topology}. Saikia et al.  propose a method for finding repeating topological structure in a scalar data using a data structure called the extended branch decomposition graph (eBDG)~\cite{saikia2014extended}. In a following paper~\cite{saikia2015fast} the authors  describe a histogram feature descriptor to compare subtrees of merge trees against each other. Narayanan et al. define a distance measure between extremum graphs  to compare two scalar fields~\cite{narayanan2015distance}. 

Many other comparison measures have been proposed in the literature for finding the distance between graphs or topological data structures. Bauer et al.  have proposed a  functional distortion metric on Reeb Graph and show its stability properties~ \cite{bauer2014measuring}.
A survey on graph edit distance by Gao et al.~\cite{gao2010survey} discusses different inexact graph matching algorithms for the application in pattern analysis. Sridharamurthy et al.  propose an edit distance between merge trees for feature visualization in  time-varying scalar data~\cite{sridharamurthyedit}. Thomas et al. propose a multiscale symmetry detection technique in scalar fields using contour clustering and studying the similarity between them ~\cite{thomas2014multiscale}.
In related works, different distance metrics between the  merge trees have been proposed to provide a similarity between the corresponding scalar fields  ~\cite{morozov2013interleaving, beketayev2014measuring}.

Other techniques that are not based on topological analysis have also been proposed in the literature for tracking and visualizing time-varying features. 
Ji et al.~\cite{Ji_featuretracking} proposed a global optimization algorithm for time-varying data and resolved the problems of volume overlapping and aggregate-attribute criteria by using the earth mover's distance. A  branch-and-bound approach was used for the global cost evaluation. The resultant approach and the metric was able to track features accurately and efficiently.
Lee et al.~\cite{Lee_TAcbased} proposed a time activity curve (TAC) to visualize time-varying features. 

However, topological feature search in time-varying multifield data is a comparatively new area of research and only few works can be found in the literature. Duke et al. \cite{2012-Duke-VisWeek} propose a joint contour net  (JCN) based visualization technique for detecting nuclear scission feature in the time-varying multifield density data. It has been observed that direct visualization of the topological features using JCNs does not scale to large data sizes because the JCN structure can be extremely complicated. In this paper, our method replaces this JCN visualization technique by a histogram comparison method.

\section{Background}
\label{sec:background}
In this section, we discuss a few tools and techniques from the literature that are required to describe our proposed distance measure.

\subsection{Histogram and isosurface statistics, continuous scatter plot}
A histogram visualizes the distribution of the samples of a scalar field using a bar graph that is constructed by binning the samples in the field range.
Histograms provide a measure of importance of isovalues based on the statistics of sample points. Carr et al.~\cite{4015490} show that histograms represent the spatial distribution of scalar fields with a nearest neighbourhood interpolation. Moreover, they show that isosurface statistics, such as the area of isosurfaces~\cite{Bajaj:1997:CS:266989.267051}, betters represent the distribution of a scalar field.

Bivariate histograms represent two fields together. These histograms consist of bins  of possibly different shapes such as square, triangle or hexagonal~\cite{scott1988note}. Square shaped bins of the histogram consist of the count for each pair of values defined on the axes. This count can be used to calculate the variance and bias from the integrated mean square error by using appropriate formulae. The square bins can be stretched to a rectangular shape based on the scale defined on the axes. 

The density function corresponding to a collection of continuous input fields is well represented by a continuous scatter plot. Unlike histograms, continuous scatter plots do not depend on the bin sizes.  
Bachthaler et al~\cite{Bachthaler:2008:CS:1477066.1477444} describe a mathematical model for generic continuous scatter plots of maps from $n$-D spatial domain to $m$-D data domain.
Lehmann et al.~\cite{Lehmann-2010} describe algorithm for detecting discontinuities in the continuous scatter plots that reveal important topological features in the data. 

\subsection{Multifield Topology and Jacobi Set}

A multifield on a $d$-manifold $\mathbb{M}\, (\subseteq\mathbb{R}^d)$ with $r$ component scalar fields
$f_i:\mathbb{M}\rightarrow \mathbb{R}$ ($i=1,\,\ldots, r$) is a \textit{map} $\mathbf{f}=(f_1,\,f_2,\,\ldots,\,f_r): \mathbb{M}\rightarrow
\mathbb{R}^r$. In differential topology, $\f$ is considered to be a \textit{smooth map} when all its partial derivatives of any order
are continuous. A point $\mathbf{x}\in \mathbb{M}$ is called a \textit{singular point} (or
\textit{critical point}) of $\mathbf{f}$ if the
rank of its differential map $\df_{\textbf{x}}$ is strictly less than $\min\{d, r\}$ where $\df_{\mathbf{x}}$ is the $r \times d$ Jacobian matrix whose rows are the
gradients of $f_1$ to $f_r$ at $\mathbf{x}$. And the corresponding value $\f(\mathbf{x})=\mathbf{c}=(c_1,\, c_2,\, \ldots,\, c_r)$ in $\mathbb{R}^r$ is a \textit{singular
value}. Otherwise if the rank of the differential map $\df_{\mathbf{x}}$ is $\min\{d, r\}$ then $\mathbf{x}$  is called a \textit{regular point} and a point $\mathbf{y}\in \mathbb{R}^r$
 is a \textit{regular value} if $\f^{-1}(\mathbf{y})$ does not contain a singular point.

The inverse image of the map $\f$ corresponding to a
value $\mathbf{c}\in \mathbb{R}^r$, $\f^{-1}(\mathbf{c})$ is called a \textit{fiber}  and each
connected component of the fiber is called a \emph{fiber-component}
\cite{Saeki2014, 2004-Saeki}. In particular, for a scalar field these are known as
the \emph{level set} and the \emph{contour}, respectively.
The inverse image of a singular value is called a \textit{singular fiber} and the inverse image
of a regular value is called a \textit{regular fiber}. If a
fiber-component passes through a singular point, it is called a \emph{singular
fiber-component}. Otherwise, it is known as a \emph{regular
fiber-component}. Note that a singular fiber may contain a regular fiber-component. Topology of a multifield data is usually studied based on its fiber-topology~\cite{2015-Chattopadhyay-CGTA-simplification}. 

Jacobi set is used to study topological relationship between two or multiple scalar fields. 
Jacobi set $\mathbb{J}_{\f}$ of a multifield $\f$ is the closure of the set of all its singular points, i.e. $\mathbb{J}_{\f}=cl\left\{\mathbf{x}\in\mathbb{M}: \mathrm{rank} ~\df_{\bx} <\min\{d, r\} \right\}$. 
Alternatively, the Jacobi set is the set of  critical
points of one component field (say $f_i$) of $\f$ restricted to the intersection of the level
sets of the remaining component fields~\cite{2004-edels-jacobi}.
 Intuitively, Jacobi set of two generic Morse functions $f_1,f_2 : \mathbb{M}\to\mathbb{R}$ is the set of points where gradients of the individual fields are parallel, i.e.
$\mathbb{J} =\{  \bx\in \mathbb{M} : \nabla{f_1(\bx)} \times \nabla{f_2(\bx)} = \mathbf{0}\}$.  Jacobi set plays a central role in the design of a comparison measure between two or multiple scalar fields~\cite{2004-edels-localglobal}.

\begin{figure*}[!h]
\centering
\subfloat[\label{fig:ht-parab}]{\includegraphics[width=3.5 cm]{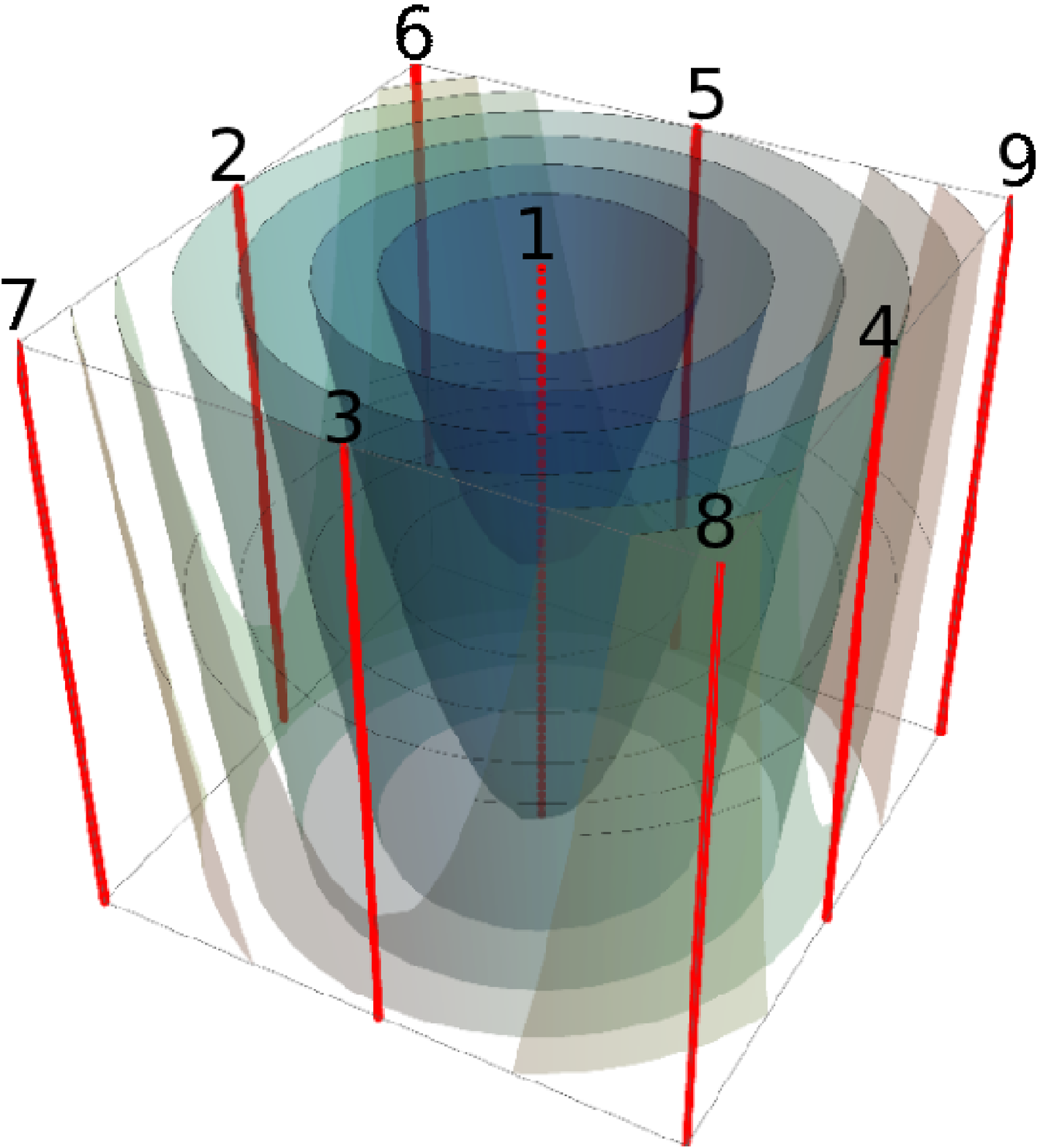}}
\subfloat[\label{fig:singular-fibers}]{\includegraphics[width=4.2cm]{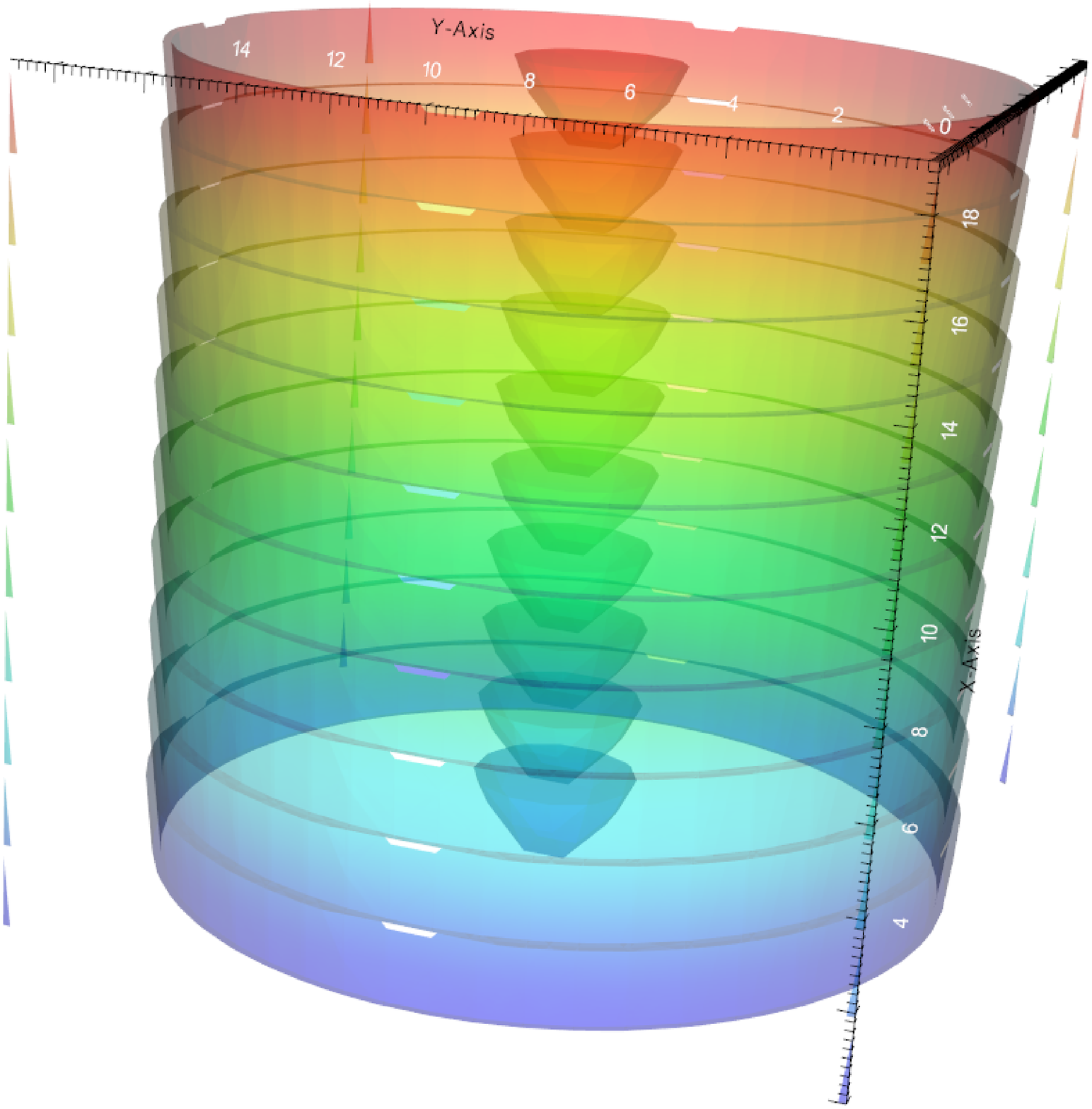}}\\
\subfloat[\label{fig:reeb-space-jacobi}]{\includegraphics[width=5cm]{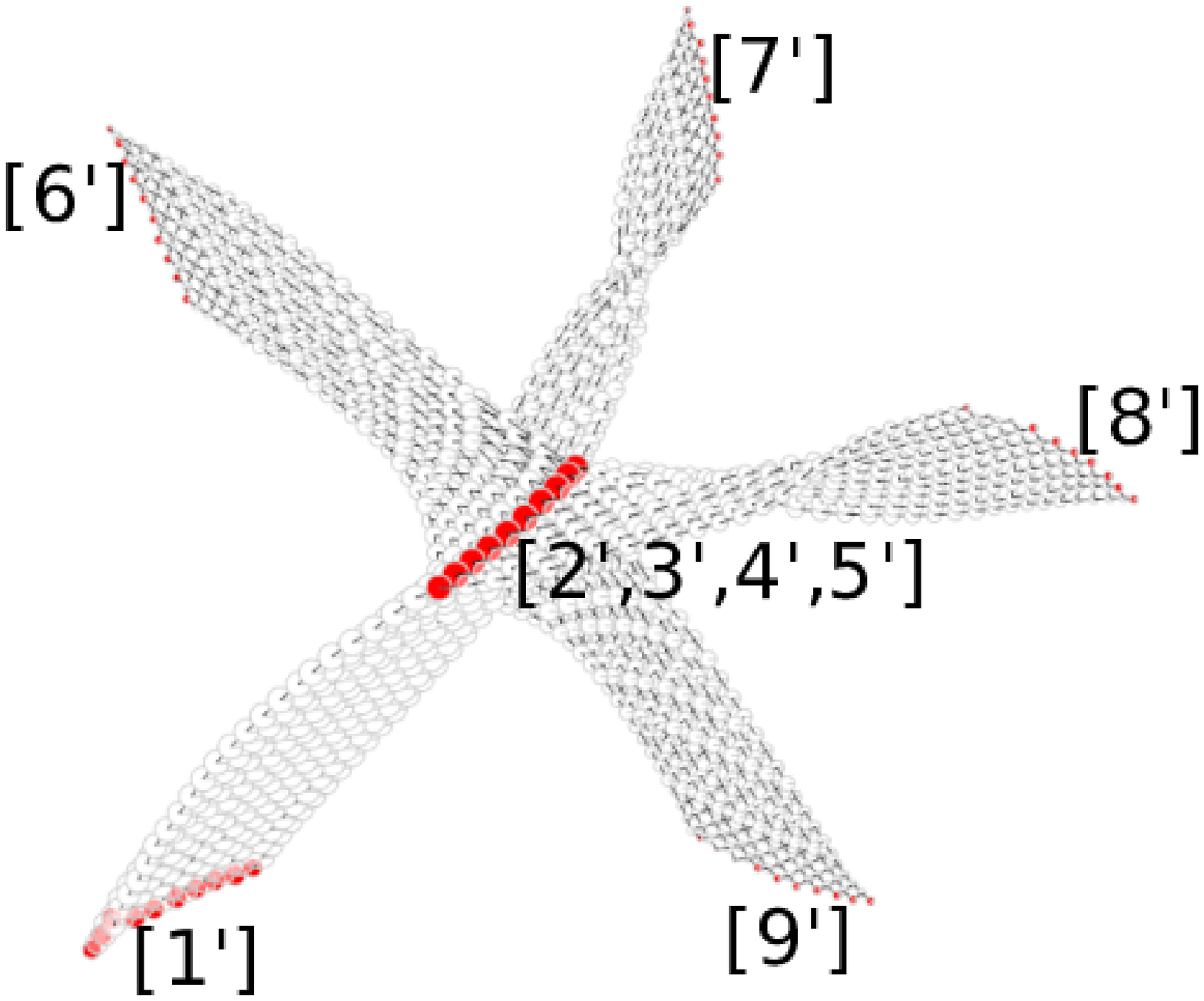}}
\subfloat[\label{fig:histogram}]{
\includegraphics[scale=0.25]{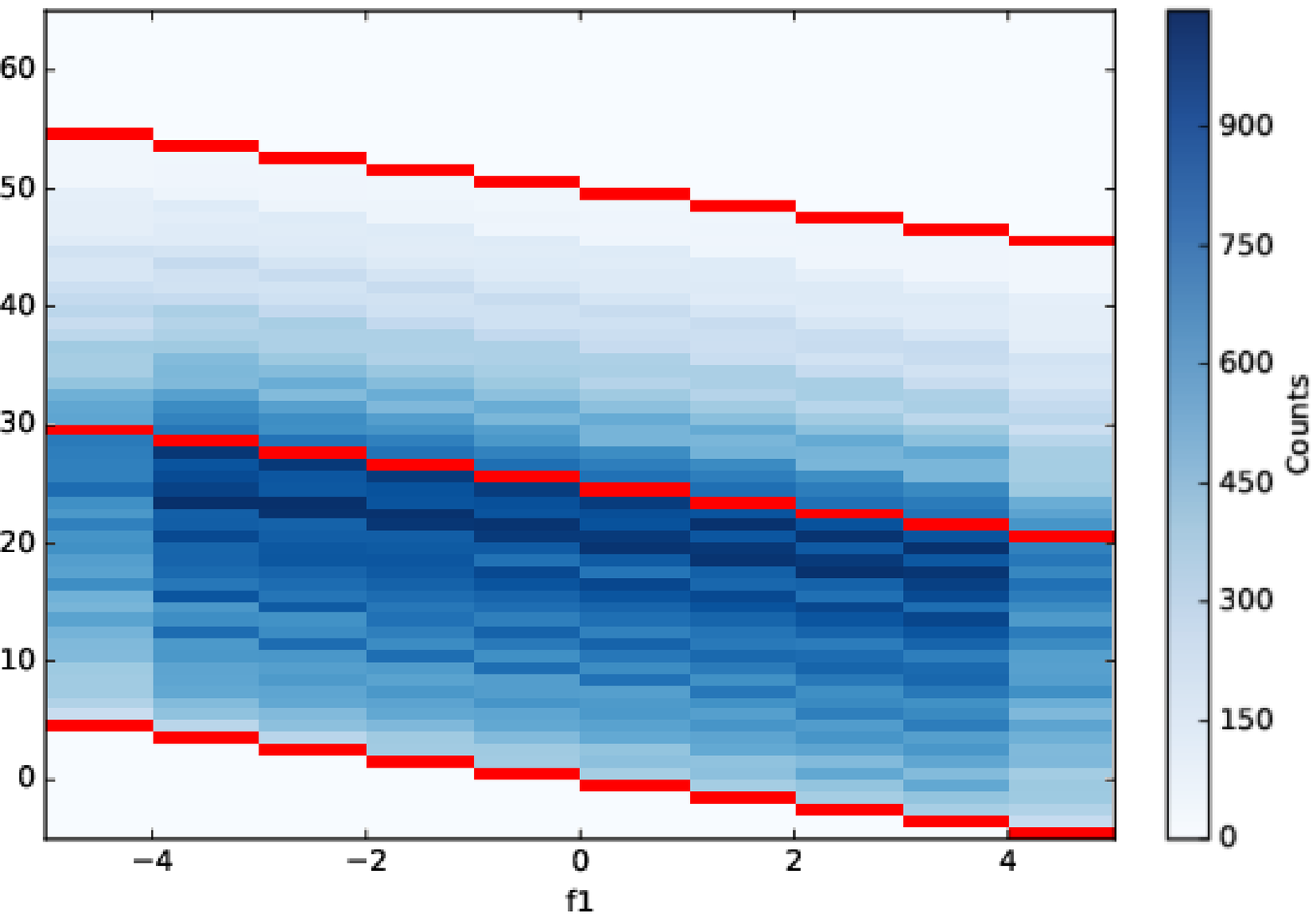}}
\caption{Figure shows a bivariate synthetic data and corresponding structures to understand its topology. (a)~Paraboloid and height field with Jacobi set (red), total $9$ connected components of the Jacobi set are numbered as $1$ to $9$ (b)~Singular fiber-components that pass through the Jacobi set points, (c)~Reeb Space (JCN) with Jacobi structure (in red). Jacobi structure components that are the projection of the Jacobi set components on the Reeb Space are shown by the corresponding dashed numbers. (d)~Histogram with singular values (bins).}
\end{figure*}
\subsection{Reeb Space and Joint Contour Net}
Similar to the Reeb graph of a scalar field,
 the Reeb space parameterizes the fiber-components of a
 multifield and its topology is described by the
 standard quotient space topology~\cite{2008-edels-reebspace}. A Jacobi structure has been defined as a projection of the Jacobi set on the Reeb space, by the quotient map ~\cite{2015-Chattopadhyay-CGTA-simplification}. 
 Figure~\ref{fig:reeb-space-jacobi} illustrates a Reeb space with Jacobi structure (in red) corresponding to a bivariate field. 

Joint Contour Net (JCN)~\cite{JCN_paper} gives a practical algorithm for approximating a Reeb space. A JCN is built in four stages. The first step of the JCN algorithm constructs all the
\emph{contour fragments} in each cell of the entire mesh  corresponding to a quantization of each component field. In the second step, the \emph{joint contour fragments} are computed by
computing the intersections of these contour fragments for the component
fields in a cell. The third step is to construct an adjacency graph (dual graph) of these
joint contour fragments where a node in the graph corresponds to a joint contour fragment and there is an edge between two nodes if the corresponding
joint contour fragments are adjacent. Finally, the JCN is obtained by collapsing the neighbouring redundant nodes with identical isovalues. Thus, each node in the JCN corresponds to a \emph{joint contour slab} or quantized fiber-component and an edge represents the adjacency between two quantized fiber-components. We use the JCN implementation for computing the quantized fiber-components and its histogram, see Figure~\ref{fig:histogram}.

\subsection{Histogram Distance Measures}\label{3.4}
Different measures have been proposed in the literature to study the  distance between two histograms~\cite{rubner2000earth}.
The measures may be classified into two types based on how they are computed -- bin-to-bin measures or cross-bin measures. In the former type, bins with the same indices are compared. We list below, a few examples of measures for finding distance between two histograms $H$ and $K$ with bin count $h_i$ and $k_i$ respectively.\\
\textbf{Minkowski-form distance: }
\begin{align}
d_{L_r}({H},{K})=\left(\displaystyle\sum_{i}|h_{i}-k_{i}|^r\right)^{1/r}
\end{align}
Commonly used Minkowski-form distances are $d_{L_1}$, $d_{L_2}$ and $d_{L_{\infty}}$. These are often used to compute dissimilarity between two color images.\\
\textbf{Histogram intersection:}
\begin{align}
d_{\cap}({H},{K})=1-\displaystyle\frac{\displaystyle\sum_{i}{\min(h_i,k_i)}}{\sum_{i}k_i}
\end{align}
This distance can capture the partial matches when the areas of the two histograms are not equal.\\
\textbf{Kullback-Leibler (KL) divergence:}
\begin{align}
d_{KL}({H},{K})=\displaystyle\sum_{i}{h_{i}\log\frac{h_i}{k_i}}
\end{align}
This is designed from an information-theoretic viewpoint. The measure is non-symmetric and sensitive to histogram binning.

One example of a cross-bin dissimilarity measure is the\\
\noindent
\textbf{Quadratic-form distance:}
\begin{align}
d_{A}({H},{K})=\displaystyle\sqrt{(\mathbf{h}-\mathbf{k})^T \mathbf{A} (\mathbf{h}-\mathbf{k})},
\end{align}
where $\mathbf{h}$ and $\mathbf{k}$ are vector representations of $H$ and $K$, respectively. The matrix $\mathbf{A}=[a_{ij}]$ is the similarity matrix
where $a_{ij}$ denote the similarity between the $i$-th bin of $H$ with the $j$-th bin of $K$~\cite{rubner2000earth}.

\section{Our Method}
\label{sec:method}
Let us consider two continuous multifields $\mathbf{f}=(X_1, X_2, \ldots, X_r)$ and $\mathbf{g}=(Y_1, Y_2, \ldots, Y_r)$ over a $d$-dimensional compact domain $\mathbb{D}\subseteq \mathbb{R}^d$ where each of $X_i$ and
$Y_i$, $(i=1,\,2, \ldots, r)$ are  real-valued scalar fields in the domain
$\mathbb{D}$. We consider comparing multifields $\mathbf{f}$ and $\mathbf{g}$ that have almost similar topological features, e.g. multifields at two consecutive time steps of a time-varying multifield data where topological features  vary continuously over time. A fiber of the multifield $\mathbf{f}$ corresponding to a
parametric point $\mathbf{c}=(c_1, c_2, \ldots, c_r)$ is the preimage $\mathbf{f}^{-1}(\mathbf{c})=X_1^{-1}(c_1)\cap X_2^{-1}(c_2)\cap \ldots \cap X_r^{-1}(c_r)$.
A connected component of the fiber is called a fiber-component.  Fiber-component topology is used to study  multifield topology, similar to the use of contour topology for scalar field studies. The Reeb space is a generalization of the Reeb graph. It captures the fiber-component topology corresponding to a multifield. However, Reeb space structure is rather complicated and computing an effective distance measure between two Reeb spaces for comparing corresponding multifields is an open problem. 

In the current work, we consider the change in fiber-component distribution over parametric space to capture the change in topology in two multifields with almost similar topological features. We observe that the change in number of fiber-components corresponding to a point on the parametric space implies the change (birth or death) in number of sheets of the Reeb Space.
Therefore, to study the topological changes from  $\mathbf{f}$ to $\mathbf{g}$ we first consider the fiber-component distributions as the feature-descriptors of the respective multifields. Next, we  propose few simple distance measures between the fiber-component distributions to capture the difference in terms of topological features.

\subsection{Fiber-Component Distribution over the Range Space}
Let $\mathbf{f}=(X_1, X_2,\ldots,X_r)$ be a continuous multifield from a $d$-dimensional compact domain $\mathbb{D}\subseteq \mathbb{R}^d$
to the $r$-dimensional range space $R_{\mathbf{f}}=[a_1,b_1]\times[a_2,b_2]\times\ldots\times[a_r,b_r]$, $a_i, b_i \in \mathbb{R}$. Define the function  $N:R_{\mathbf{f}}\rightarrow\mathbb{N}$ as  $N(\bx)=|\mathbf{f}^{-1}(\bx)|$ for $\bx \in R_{\mathbf{f}}$, where $|\mathbf{f}^{-1}(\bx)|$ represents the number of connected components in the fiber $\mathbf{f}^{-1}(\bx)$.  In other words, $N(\bx)$ maps each point $\bx$ of $R_{\mathbf{f}}$ to the corresponding number of fiber-components of $\textbf{f}$. We assume that $N$ is a bounded function for multifields $\mathbf{f}$ defined over a compact domain $\mathbb{D}$. To compute the total number of fiber-components, we partition the range $R_{\mathbf{f}}$ into a union of $m^r$ sub-boxes by introducing the partitions of the intervals: $a_i=x_{0}^{(i)}<x_{1}^{(i)}<\ldots<x_{m}^{(i)}=b_i$ for $i=1,2, \ldots,r$. Let $\bx_{i_1i_2\ldots i_r}$ be a point in the sub-box $B_{i_1i_2\ldots i_r}=[x_{i_1-1}^{(1)},x_{i_1}^{(1)}]\times [x_{i_2-1}^{(2)},x_{i_2}^{(2)}]\times\ldots \times [x_{i_r-1}^{(r)},x_{i_r}^{(r)}]$ for $i_1, i_2, \ldots, i_r=1, 2, \ldots, m$ with volume $\Delta V_{i_1i_2\ldots i_r}$. Then, $\mathbf{N}$, defined as the sum of number of fiber-components over all points in $R_{\mathbf{f}}$  is equal to 
\begin{align}
\mathbf{N}=\lim_{\text{all }\Delta V_{i_1i_2\ldots i_r}\rightarrow 0}\sum_{i_1, i_2, \ldots, i_r=1}^m N(\bx_{i_1i_2\ldots i_r})\Delta V_{i_1i_2\ldots i_r}=\displaystyle\int_{R_{\mathbf{f}}}{N(\bx) d\mathbf{x}}.
\end{align} 
The function $N$ is bounded and hence integrable.
Next, we define a density function of the fiber-component distribution as:

\begin{align}
\label{eq:pdf-noofcomp-cont}
\mathfrak{p}_{\mathbf{f}}(\mathbf{x})=\frac{N(\mathbf{x})}{\mathbf{N}} \text{ for } \mathbf{x}\in R_{\mathbf{f}},
\end{align}

\noindent
where $$\displaystyle\int_{R_{\mathbf{f}}} \mathfrak{p}_{\mathbf{f}}(\mathbf{x})d\mathbf{x} =1.$$

In practice, to compute the fiber-component distribution over the range space, we first discretize the continuous multifield $\mathbf{f}=(X_1, X_2,\ldots,X_r)$ 
in the $r$-dimensional range space. Let field $X_i$ be discretized (quantized) uniformly at the values $x_{1}^{(i)}< x_{2}^{(i)}< \ldots <x_{m_i}^{(i)}$ for $i=1, 2, \ldots, r$.
 We denote this discrete range space as $\mathrm{spec}(R_{\mathbf{f}})=I_1\times I_2\times \ldots \times I_r$, the Cartesian product of $I_i=\{x_{1}^{(i)},\,x_{2}^{(i)},\, \ldots,\, x_{m_i}^{(i)}\}$ ($i=1, 2,\ldots, r$). Then we compute the frequency distribution of the corresponding fiber-components over this discrete range space (spectrum). The probability mass function of the corresponding discrete probability distribution is given by
\begin{align}
\label{eq:pdf-noofcomp}
p_{\mathbf{f}}({\mathbf{x}})=\frac{\tilde{N}_{\mathbf{x}}}{\tilde{\mathbf{N}}},\;\text{where}\; \mathbf{x}\in \mathrm{spec}(R_{\mathbf{f}}).
\end{align}

\noindent
Here, $\tilde{N}_{\mathbf{x}}$ counts the number of fiber-components at the parametric point  $\mathbf{x}=(x_{i_1}^{(1)}, x_{i_2}^{(2)}, \ldots, x_{i_r}^{(r)})$ in $\mathrm{spec}(R_{\mathbf{f}})$ (for $i_1=1, 2, \ldots, m_1$; $i_2=1, 2, \ldots, m_2$; \ldots; $i_r=1, 2, \ldots, m_r$) and $\tilde{\mathbf{N}}$ is the sum of
number of fiber-components of $\mathbf{f}$ over all points in the discrete range space $\mathrm{spec}(R_{\mathbf{f}})$. Note that $p_{\mathbf{f}}$ defines a probability mass function (p.m.f.) since
$p_{\mathbf{f}}(\mathbf{x})\geq 0$ and 
$$\displaystyle\sum_{\mathbf{x}\in \mathrm{spec}(R_{\mathbf{f}})}p_{\mathbf{f}}(\mathbf{x})=1.$$
\noindent
When the quantization level goes to infinity then discrete case converges to the continuous case.
Alternatively, one can define p.m.f. using $A_{\mathbf{x}}$ by measuring the size of the quantized fiber-components at the parametric point $\mathbf{x}\in \mathrm{spec}(R_{\mathbf{f}})$ and $A$ is the total
measure of all the fiber-components over $\mathrm{spec}(R_{\mathbf{f}})$. Thus we have 
\begin{align}
\label{eq:pdf-volume}
p_{\mathbf{f}}({\mathbf{x})=\frac{A_{\mathbf{x}}}{A},\;\text{where}\; \mathbf{x}\in \mathrm{spec}(R_{\mathbf{f}})}.
\end{align}
In the proposed distance measure that we will describe next, we consider the definitions in (\ref{eq:pdf-noofcomp-cont}) and (\ref{eq:pdf-noofcomp}) because they capture the topological changes in the fibers of the multifield.

\subsection{Distance between two Fiber-Component Distributions}\label{4.2} 
Let us consider two multifields $\mathbf{f}_1=(X_1, X_2, \ldots, X_r)$ and $\mathbf{f}_2=(Y_1, Y_2, \ldots, Y_r)$ over the domain $\mathbb{D}\subseteq \mathbb{R}^d$. Let $R_{\mathbf{f}_1}$ and $R_{\mathbf{f}_2}$ be the range spaces of $\mathbf{f}_1$ and $\mathbf{f}_2$, respectively. We note that the range spaces $R_{\mathbf{f}_1}$ and $R_{\mathbf{f}_2}$ may be different but restrict our attention to the case when they are almost equal. To define our distance measures between the fiber-component distributions of $\mathbf{f}_1$ and $\mathbf{f}_2$, first we extend the range spaces $R_{\mathbf{f}_1}$ and $R_{\mathbf{f}_2}$ to an equal range $R$. We define $R$ as: $R=R_1\times R_2\times \ldots \times R_r$ where $R_i=\mathrm{range}~X_i \cup \mathrm{range}~Y_i$  for $i=1, 2, \ldots, r$. This extended range $R$ is considered as the common domain of fiber-component distributions of both $\mathbf{f}_1$ and $\mathbf{f}_2$. The fiber-component distributions of $\mathbf{f}_1$ on the part $R\setminus R_{\mathbf{f}_1}$, corresponding to which $\mathbf{f}_1$  has no data, is filled with zeros. Similarly fiber-component distributions of $\mathbf{f}_2$ on $R\setminus R_{\mathbf{f}_2}$ is filled with zeros.

For the continuous case: let $\mathfrak{p}_{\mathbf{f}_1}$ and $\mathfrak{p}_{\mathbf{f}_2}$ be the density functions of the fiber-component distributions of $\mathbf{f}_1$ and $\mathbf{f}_2$, respectively, over the extended range $R$. Let $\mathbf{P}_1$ and $\mathbf{P}_2$ be the corresponding distribution functions. Then we define a point-wise distance measure between $\mathbf{P}_1$ and $\mathbf{P}_2$ as:
\begin{align}
d_q(\mathbf{P}_1,\mathbf{P}_2)=\left(\displaystyle\int_R|\mathfrak{p}_{\mathbf{f}_1}(\mathbf{x})-\mathfrak{p}_{\mathbf{f}_2}(\mathbf{x})|^qd\mathbf{x}\right)^{1/q}
\end{align}
\noindent 
for any real number $q\geq 1$. In particular for $q=1$, $q=2$ or $q=\infty$ we get similar distance measures of practical importance.

For the discrete case, let the range space $R$ be discretized (quantized) as
$\mathrm{spec}(R)=I_1\times I_2\times \ldots \times I_r$ where $I_i=\{x_{1}^{(i)},\,x_{2}^{(i)},\, \ldots,\, x_{m_i}^{(i)}\}$.
Let $\mathbf{P}_1=\{p_{\mathbf{x}}^{(1)}:\; \mathbf{x}\in \mathrm{spec}(R)\}$ and $\mathbf{P}_2=\{p_{\mathbf{x}}^{(2)}:\; \mathbf{x}\in \mathrm{spec}(R)\}$ be the fiber-component distributions of $\mathbf{f}_1$ and $\mathbf{f}_2$, respectively, over the discrete range space $\mathrm{spec}(R)$.
Then we define the point-wise distance measure between  the distributions $\mathbf{P}_1$ and $\mathbf{P}_2$ as:
\begin{align}
d_q(\mathbf{P}_1,\mathbf{P}_2)=\left(\displaystyle\sum_{\mathbf{x}\in\mathrm{spec}(R)}|p^{(1)}_{\mathbf{x}}-p^{(2)}_{\mathbf{x}}|^q\right)^{1/q}.
\end{align}
for any real number $q\geq 1$. In particular, for $q=1$, $q=2$ and $q=\infty$ we have

\begin{align}
d_1(\mathbf{P}_1,\mathbf{P}_2)=\displaystyle\sum_{\mathbf{x}\in\mathrm{spec}(R)}|p^{(1)}_{\mathbf{x}}-p^{(2)}_{\mathbf{x}}|
\end{align}

\begin{align}
d_2(\mathbf{P}_1,\mathbf{P}_2)=\left(\displaystyle\sum_{\mathbf{x}\in\mathrm{spec}(R)}|p^{(1)}_{\mathbf{x}}-p^{(2)}_{\mathbf{x}}|^2\right)^{1/2}
\end{align}
\noindent
and
\begin{align}
d_{\infty}(\mathbf{P}_1,\mathbf{P}_2)=\sup_{\mathbf{x}\in \mathrm{spec}(R)}|p^{(1)}_{\mathbf{x}}-p^{(2)}_{\mathbf{x}}|.
\end{align}

\noindent
These distance measures are motivated from the observation that the point-wise difference $|\tilde{N}_{\mathbf{x}}^{(1)}-\tilde{N}_{\mathbf{x}}^{(2)}|$ captures the number of changes in fiber-components between two multifields at consecutive time steps for $\mathbf{x}\in \mathrm{spec}(R)$. Note that each fiber-component of a multifield corresponds to exactly one sheet of its Reeb space. So, the difference in number of fiber-components captures the number of possible changes in Reeb space sheets containing the parameter value $\mathbf{x}$. Thus, $|\tilde{N}_{\mathbf{x}}^{(1)}-\tilde{N}_{\mathbf{x}}^{(2)}|$ captures the number of births or deaths of sheets containing the parameter value $\mathbf{x}$ of the corresponding Reeb spaces. 

\subsection{Weighted Distance for the Singular Values}\label{4.3}
Singular fibers capture  the topological changes in the evolution of fibers in a multifield. The image of a singular fiber in the parametric space is called a singular value. Because of importance of the singular values compare to regular values, we propose a variant to the distance measure that weights the singular values differently,
\begin{align}
\label{eqn:metric-singular}
d_q^{\mathbb{S}}(\mathbf{P}_1,\mathbf{P}_2; \omega)= \left[\omega\displaystyle\sum_{\mathbf{x}\in \mathbb{S}}|p^{(1)}_{\mathbf{x}}-p^{(2)}_{\mathbf{x}}|^q +\displaystyle\sum_{\mathbf{x}\notin \mathbb{S}}|p^{(1)}_{\mathbf{x}}-p^{(2)}_{\mathbf{x}}|^q\right]^{1/q}.
\end{align}

\noindent
Here, $\mathbb{S}$ is the set of singular values in the discrete range space $\mathrm{spec}(R)$ and $q\geq 1$. Moreover, $\omega > 1$ is the weight parameter to impose more importance to the singular values than the regular values. We observe from our experiments on different  datasets that increasing the weight $\omega$ increases the prominence of the events that correspond to topological changes when we plot weighted distances over time. 
Figure~\ref{fig:histogram} shows a fiber-component histogram with the singular values (in red) corresponding to the bivariate field in Figure~\ref{fig:ht-parab}.

\subsection{Metric Space Properties of the Distance Measures}\label{4.4}
It is important to show that the proposed distance measures between two distributions satisfy the metric space properties for the space $\mathcal{P}_R$ of all possible fiber-component distributions corresponding to different multifields with range $R$. Let us first show that $(\mathcal{P}_R, d_q)$ is a  metric space.
\begin{enumerate}
\item \textbf{Non-negativity.} Note $d_q$ is real-valued, finite and non-negative.

\item \textbf{Identity.} We note that for two distributions $\mathbf{P}_1, \mathbf{P}_2\in \mathcal{P}_R$, $d_q(\mathbf{P}_1, \mathbf{P}_2)=0$ if and only if $\mathbf{P}_1=\mathbf{P}_2$, since $\displaystyle\displaystyle\sum_{\mathbf{x}\in\mathrm{spec}(R)}|p^{(1)}_{\mathbf{x}}-p^{(2)}_{\mathbf{x}}|^q=0$ implies $p^{(1)}_{\mathbf{x}}=p^{(2)}_{\mathbf{x}}$ for all $\mathbf{x}\in \mathrm{spec}(R)$.

\item \textbf{Symmetry.} It is straight-forward to show that $d_q(\mathbf{P}_1, \mathbf{P}_2)=d_q(\mathbf{P}_2, \mathbf{P}_1)$. This implies the symmetry property of $d_q$.

\item  \textbf{Triangle inequality.} To show the triangle inequality of $d_q$ we consider three fiber-component distributions $\mathbf{P}_1$, $\mathbf{P}_2$ and $\mathbf{P}_3$. Note, for $q=1$, $|p^{(1)}_{\mathbf{x}}-p^{(3)}_{\mathbf{x}}|\leq |p^{(1)}_{\mathbf{x}}-p^{(2)}_{\mathbf{x}}|+|p^{(2)}_{\mathbf{x}}-p^{(3)}_{\mathbf{x}}|$. For $q\geq 1$, using Minkowski inequality~\cite{Hardy-1952} we can show that $d_q(\mathbf{P}_1, \mathbf{P}_3)\leq d_q(\mathbf{P}_1, \mathbf{P}_2)+d_q(\mathbf{P}_2, \mathbf{P}_3)$.
\end{enumerate}

\noindent
Similar properties can be proved for the other distance measures $d^{\mathbb{S}}_{q}$, $d_1$, $d_2$ and $d_{\infty}$.
However, note the above metric properties hold in the space of fiber-component distributions, not necessarily in the  space of actual multifields.


\section{Implementation}
\label{sec:implementation}
We implement the distance measures described in the previous section using Visualization Toolkit (VTK) \cite{kitware2003} under the Joint Contour Net \cite{JCN_paper} implementation framework. The implementation works for a generic pair for multifields but is particularly designed for time-varying multifields. We note that the range spaces of two multifields at two consecutive time steps are not necessarily the same and may vary slightly. We expand the range of both multifields by considering their component wise union and use zero-padding to compute the histogram as described in section~\ref{4.2}. Next, we describe the four main steps of our implementation.

\noindent
I. \textbf{Computing Fiber-Components:} First, we discretize or quantize the common range of the multifields into finite numbers of bins. Then corresponding to each bin-value, we compute the quantized fiber-components as described in the JCN algorithm~\cite{JCN_paper}. 
In other words, compute the contour slabs in each cell for each of the scalar fields and then find intersection of the slabs to get the fragments. Finally an adjacency graph is computed from the fragments to obtain quantized fiber-components. Each quantized fiber-component corresponds to a node of the JCN.

\noindent
II. \textbf{Computing Fiber-Component Histograms:}
Next, we compute the  $r$-dimensional fiber-component histogram corresponding to each multifield on the range space. We use the same binning as used for the quantized fiber-component computation.  Each bin in the range is populated with the corresponding fiber-components. We compute the number of fiber-components in each bin for the fiber-component histogram computation.
 A color map specifying the number of all the nodes is shown in Figure~\ref{fig:histogram}. The color map is chosen over a range of blue values. Light blue shows fewer number of nodes (fiber-components), and as the color darkens the number of nodes (fiber-components) also increases. 

\noindent
III. \textbf{Computing Singular Values of Multifields:}
To compute singular values first one needs to compute the singular points or the Jacobi set in the domain of the multifield and then the corresponding range values of those points are actually the singular values. In the current implementation we first compute the Jacobi structure using a multi-dimensional Reeb graph (MDRG) as described in~\cite{2015-Chattopadhyay-CGTA-simplification, 2014-EuroVis-short} and then project them in the histogram-bins and call those bins as singular bins.
We note that a singular bin of the histogram may contain both singular and regular fiber-components (nodes). In the histogram plot~Figure~\ref{fig:histogram}, the red colored bins indicate the singular bins and blue are the regular bins. For the singular bins of the histogram the singular, regular and total nodes (singular and regular together) are stored separately for further computation. 

\noindent
IV. \textbf{Computing Distance Metrics between Histograms:}
The above three steps are performed for multifields at all the time stamps or sites, and the corresponding histograms are stored in different files. A python script is then implemented to compute the corresponding probability density from the histogram.  Then  the distance metrics between two probability densities at the consecutive time steps are computed as in sections~\ref{4.2} and \ref{4.3}. The distance metric $d_q^{\mathbb{S}}$ (as in equation~\ref{eqn:metric-singular}) is computed for different values of $q$ and $\omega$. This metric is computed using the singular and regular nodes. Note that if $q=1$  and $\omega = 1$ the metric $d_q^{\mathbb{S}}$ is same as $d_1$. To validate the experiment $d_1$ is calculated using all the nodes (regular and singular nodes together).
Along with the measures that we have proposed we even calculated the distance measures for the already defined metrics for histogram comparison as defined in section~\ref{3.4}. The values for these distance metrics are stored and then used to create a comparison line plot. The values were also used to check the metric properties defined in section~\ref{4.4}. We also calculated the simple root mean square distance for bivariate data for experimental comparison.
%


\section{Applications}
\label{sec:application}
We now describe applications of the proposed comparison driven feature search method to four different datasets, namely (i) a synthetic data consisting of two polynomial functions, (ii) the scission data of plutonium atom, (iii) fission data of Fermium atom and (iv) the DFT data of carbon monoxide and platinum (CO-Pt) molecular bond. 

\subsection{Synthetic Data}
\begin{figure*}[!h]
\centering
\includegraphics[width=\textwidth]{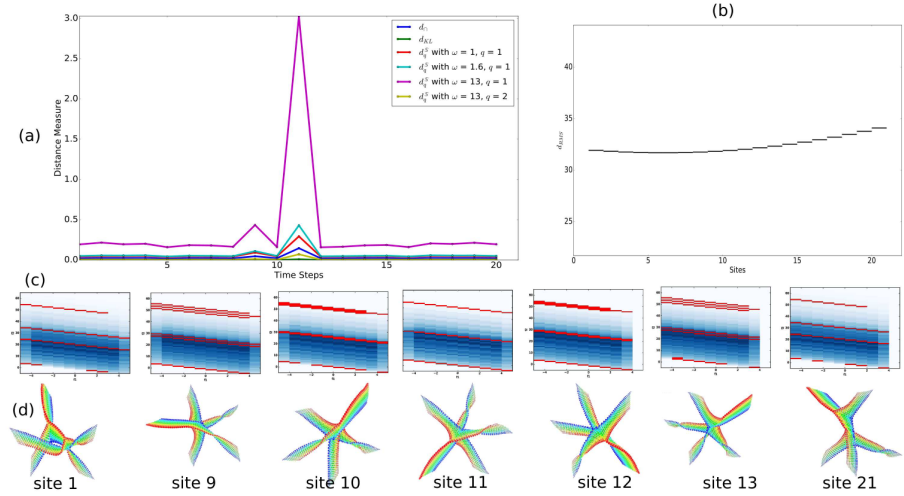}
\caption{Plots of distance measures between consecutive sites in a series of bivariate (height, paraboloid) fields. (a)~Various distance measures show a peak at site 11, indicating a topological change. The proposed metric $d_q^{\mathbb{S}}$ also exhibits a peak, more significant than other distance measures.(b)~Root-mean-square plot is not able to capture the topological change. This indicates the need for a topological data structures for multifield data that captures topological changes. (c)~Fiber-component distributions for selected sites. Singular values are highlighted in red. Blue nodes indicate regular nodes and the shades of blue indicate the number of nodes in a particular bin (light indicates low).  (d)~Corresponding Reeb spaces. The height field is mapped to color (blue is low and red is high). }
\label{fig:synthetic-plots}
\end{figure*}
We generate a synthetic bivariate field whose components are the height field $f_1(x,y,z) = z$ and the paraboloid field $f_2(x,y,z) = x^2 + y^2 -z$. Both fields are defined on an axis-aligned box $[-5.5,\,4.5]\times[-5.5,\,4.5]\times[-5.5,\,4.5]$ and sampled on a grid of size $20 \times 20 \times 20$. Next, we generate a sequence of multifield data by  incrementally translating the domain-box  along each of the three axes with small magnitude $0.05$, i.e. if $(C_x,C_y,C_z)$ and $(c_x,c_y,c_z)$ are respectively the coordinates of a point on the box before and after the translation, then $C_x=c_x+0.05,\, C_x=c_y+0.05,\, C_z=c_z+0.05$. In total, we create $21$ bivariate datasets. To create the consecutive datasets, we begin with the domain $[-5.5,\,4.5]\times[-5.5,\,4.5]\times[-5.5,\,4.5]$ and then apply the above described sequence of translations $21$ times until we obtain the domain of the final dataset, namely  $[-4.5,\,5.5]\times[-4.5,\,5.5]\times[-4.5,\,5.5]$.
The major topological feature is expected in the dataset corresponding to the domain $[-5,\, 5]\times[-5,\, 5]\times [-5,\, 5]$ (which is symmetric about origin) because of  degenerate intersections of the fiber-components with the boundary of the box. 

\subsubsection*{Observations and Results}
We compute the fiber-component histograms for each dataset in the series and plot the distance between two consecutive datasets, see  Figure~\ref{fig:synthetic-plots}. The distance peaks at site $11$ as expected. 
The red color in the histograms indicates singular nodes and blue color indicates regular nodes. The number of regular nodes in a particular bin is mapped to different shades of blue. Colors in the Reeb space indicate the height field value. Although various distance measures are able to capture the topological change, the peak was not sharp enough. The peak is most prominent using the $d_q^\mathbb{S}$ metric and increased weight for singular nodes.
Note that all the subsequent experiments are done with $\omega = 13$ in order to keep the consistency in our experiments for all the datasets. If the value of $\omega$ is increased better peaks can be obtained and the value is not dependent on the chosen dataset.

\subsubsection*{Comparison with the Root Mean Squares Metric}
To show the usefulness of the proposed metrics, we compute the distance between two multifields by directly extending the root mean square metric. The root mean square distance between two multifields $\mathbf{f}=(f_1, \ldots, f_r)$ and $\mathbf{g}=(g_1,\ldots, g_r)$ can be generalized as the square root of the mean of the sum of the difference between consecutive component fields: 
\begin{align*}\label{eq13}
d_{RMS} = \sqrt{\dfrac{1}{m}\displaystyle\sum_{i=1}^m\{(f_1(x_i)-g_1(x_i))^2 +\cdots + (f_r(x_i)-g_r(x_i))^2\}}.
\end{align*}
Here $m$ is the number of data points in the domain.  Figure~\ref{fig:synthetic-plots}(b) shows the root mean square distance metric plot. We  observe that the rms metric is not capable of capturing the topological change. This further motivates the study of measures such as the one proposed in this paper for comparing multifield data.

\subsection{Plutonium Atom Dataset}
\begin{figure*}[!h]
\centering
\includegraphics[width=\textwidth]{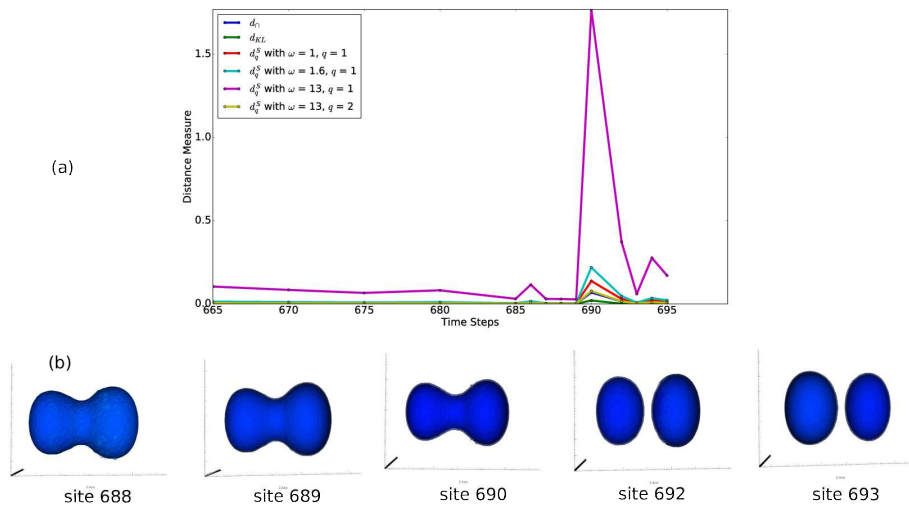}
\caption{Plots of the distance measures for the scission data for the plutonium atom. (a)~Distance measure between fields at consecutive time steps vs. the time step in the range [$665-699$].   The proposed distance measure \textbf{$d_q^{\mathbb{S}}$} exhibits a prominent peak between time step $690-692$, which indicates a significant change. (b)~Geometry of the plutonium atom at various time steps. The point of scission is between site $690-692$ and can be seen in the geometry.}
\label{fig:plutonium}
\end{figure*}
Nuclear Density Functional Theory (DFT) is an approach to understand the nuclear fission occurring in a nucleon-nucleon interaction in atomic nuclei. Nuclear fission is a process by which an atom's nucleus splits into two or more fragments. The splitting of the nucleus can be identified as stretching the core, hence it involves some deformation. This deformation can be a crucial indicator of the topology of the atom's nucleus. An important problem in nuclear fission study is the accurate identification of points in a continuous high dimensional manifold where the core  is split. The time when the atom breaks into multiple fragments is known as nuclear scission. At this time the topology of the atom changes in terms of the number of components. Physicists typically identify this phenomenon via tedious manual process. Previous works have described a visual approach to identification of scission~\cite{2012-Duke-VisWeek}. However, these methods require the inspection of the geometry of the Reeb space for all time steps. Further, the Reeb space is a complex structure that is difficult to examine. We aim to detect the key time steps that correspond to topological changes by plotting a graph of the distance measure over time.

The dataset consists of nuclear densities of plutonium atom which represents the internal structure of a heavy nucleus. The dataset is a multifield data consisting of spatial density of proton, the spatial density of neutrons and spatial density of nucleons (protons + neutrons) in the nucleus. These densities, represented as \textbf{p}, \textbf{n} and  \textbf{t} are sampled on a $40\times40\times66$ grid. The dataset available to us is a negative log transformed sample at 14 different time steps, namely [665, 670, 675, 680, 686, 687, 688, 689, 690, 692, 693, 694, 695, 699]. The time step where the nuclear scission occurs is reported in earlier work~\cite{2012-Duke-VisWeek} and confirmed by physicists.  
We use sufficiently small slab width to capture the topological change. 
We use the following parameters in our experiments:
\textbf{p} (slab width 8) and \textbf{n} (slab width 2),
\textbf{p} (slab width 8) and \textbf{t} (slab width 2),
\textbf{n} (slab width 2) and \textbf{t} (slab width 2).

\subsubsection*{Observations and Results}
We experiment with  all combination of proton, neutrons and nucleon density considering two fields at a time. The plots in Figure~\ref{fig:plutonium} show the distance measure for the first combination, \textbf{p} (slab width 8) and \textbf{n} (slab width 2).
We observe a sudden change between time steps 690 and 692. The $d_1$ distance was typically in the range of 0.0 to 0.02, but at nuclear scission, the measure increases to 0.1. This is due to the change in the number of quantized fiber-components in the range space.  After scission, the distance measure dropped down to small values because the number of fiber-components does not change after the split. Figure~\ref{fig:plutonium}(a) shows a comparison with other bin-to-bin measures that are also able to capture the topology change but the peak is not as prominent. We plot the measure $d_{q}^\mathbb{S}$ for different values of $q$ and weights. As the weight for singular values is increased, the peak becomes more prominent and as $q$ is increased the plot becomes smoother. Figure~\ref{fig:plutonium} shows the highest peak in the plot using weight $\omega=13$ (for singular bins) and $q=1$.

\subsection{Fermium Atom Dataset}
\begin{figure*}[!h]
\centering
\includegraphics[width=\textwidth]{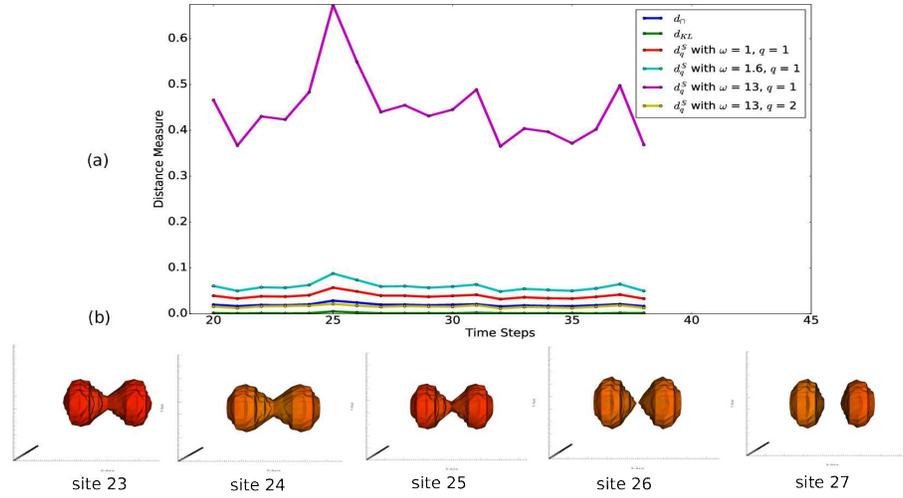}
\caption{Plots of the distance measures for the scission data for the fermium-256 atom. (a)~Distance measure between fields at consecutive time steps vs. the time step in the range [$20,39$]. The proposed distance measure \textbf{$d_q^{\mathbb{S}}$} exhibits a prominent peak at time step $26$, which indicates a significant change. (b)~Geometry of the fermium-256 atom at various time steps. The point of scission is at site $26$ and can be seen in the geometry.}
\label{fig:fermium}
\end{figure*}
We experiment with another scission dataset, namely that of the Fermium-256 atom. In this dataset, our goal is again to find the point where nuclear scission occurs. As described in  the literature~\cite{2012-Duke-VisWeek}, this dataset consists of  three different types of data viz. aEF: asymmetric elongated fission,  sCF: symmetric compact fission and sEF: symmetric elongated fission. The  dataset  that  was  made available is the sCF data and was sufficient to detect the topological change where the fermium nucleus scission happens symmetrically. The sCF dataset consists of three fields i.e. proton density (\textbf{p}), neutron density (\textbf{n}) and total density (\textbf{t}) defined on a $19\times19\times19$ sized grid. The field is available at 56 regularly spaced time steps.  Time steps 20-55 were chosen for analysis. Choosing the slab width was still an issue, and we end up working with the same slab width as that for Plutonium atom data, namely
\textbf{p} (slab width 8) and \textbf{n} (slab width 2),
\textbf{p} (slab width 8) and \textbf{t} (slab width 2),
\textbf{n} (slab width 2) and \textbf{t} (slab width 2).
\subsubsection*{Observations and Results}
The same set of experiments were done using the fermium-$256$ atom dataset. Figure~\ref{fig:fermium} shows the plots with  proton and neutron density data from time step $20$ to $39$. We observe a topological change at time step $26$. Other bin-to-bin histogram metrics, e.g. the KL divergence and the histogram intersection,  exhibit a much smaller peak as compared to the proposed $d_q^\mathbb{S}$ distance.

\subsection{Chemistry Data: Pt-CO Bond}
\begin{figure*}[!h]
\centering
\includegraphics[width=\textwidth]{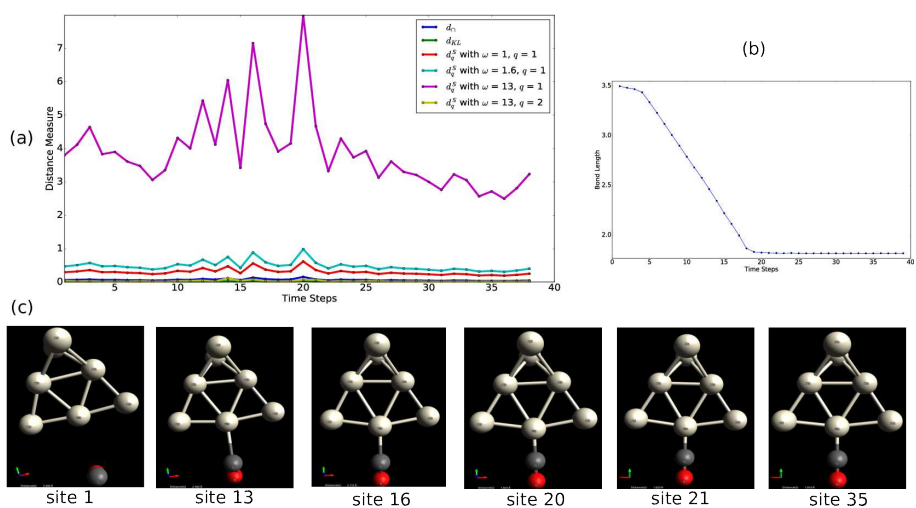}
\caption{Plots of the distance measures for the orbital density data of Pt-CO bond at different time steps. (a)~Distance measure between fields at consecutive time steps vs. the time step in the range [$0,39$]. The plots are for two field values, HOMO and LUMO and the highest peak is obtained at time stamp 21. The proposed distance measure \textbf{$d_q^{\mathbb{S}}$} exhibits a prominent peak, which indicates a significant change. (b)~Pt-CO Bond length vs time. Bond length stabilizes at time step 21. (c)~Geometry of the Pt-CO bond creation at various time steps, visualized using the tool Avogadro. Although the bond is visible at time step 13, the bond length is not stable at this site.}
\label{fig:chemistry}
\end{figure*}
Adsorption of gas molecules on metal surfaces has various applications including heterogeneous catalysis, electrochemistry, corrosion, and molecular electronics~\cite{book-Surface-Chem, 2010-JACS-Kendric}. Particularly, the adsorption of the CO molecule on platinum surfaces has attracted attention of a wide scientific community, due to its role in the areas of automobile emission, fuel cells and other catalytic processes~\cite{2003-Kresse-Pt-CO, 1995-Gasteiger-H2-CO}. Therefore, an atomic-level understanding of the CO molecule interacting with the Pt surface is of utmost importance. In this study, we have considered seven Pt atoms representing a platinum surface which interacts with a CO molecule. As the CO molecule approaches towards one of the Pt atoms, the CO bond starts weakening, and Pt-CO bond formation takes place.  Quantum mechanical computations were used to generate the electron density distribution corresponding to the highest occupied molecule orbital (HOMO), lowest occupied molecular orbital (LUMO) and HOMO$-1$. The electron density distribution was computed for varying distance between the carbon atom of the CO molecule and the Pt atom. The Pt-CO bond forms when the distance between the Pt atom and the CO molecule becomes $\sim 1.83A$.  
This Pt-CO dataset consists of orbital density for orbital numbers $69$, $70$ and $71$. Orbital number $70$ corresponds to HOMO, orbital number $71$ to LUMO and orbital number $69$ to HOMO$-1$.

\subsubsection*{Observations and Results}
Figure~\ref{fig:chemistry} shows different plots for the Pt-CO dataset.  At site $21$,  we get the most stable bond length between Pt and CO molecule. We observe that although the bond is formed at site $13$ (as validated by the geometry), the bond-length is not stable. The bond length stabilizes at site $21$ and does not change much later.
We observe a sharp peak in the plot of the proposed $d_q^\mathbb{S}$ distance. This peak corresponds to the formation of the stable bond.

\section{Single Scalar Field vs. Multifield}
\label{sec:scalar-multi}
\begin{figure*}[h!]%
    \begin{center}
    \subfloat[]{{\includegraphics[width=6.2cm]{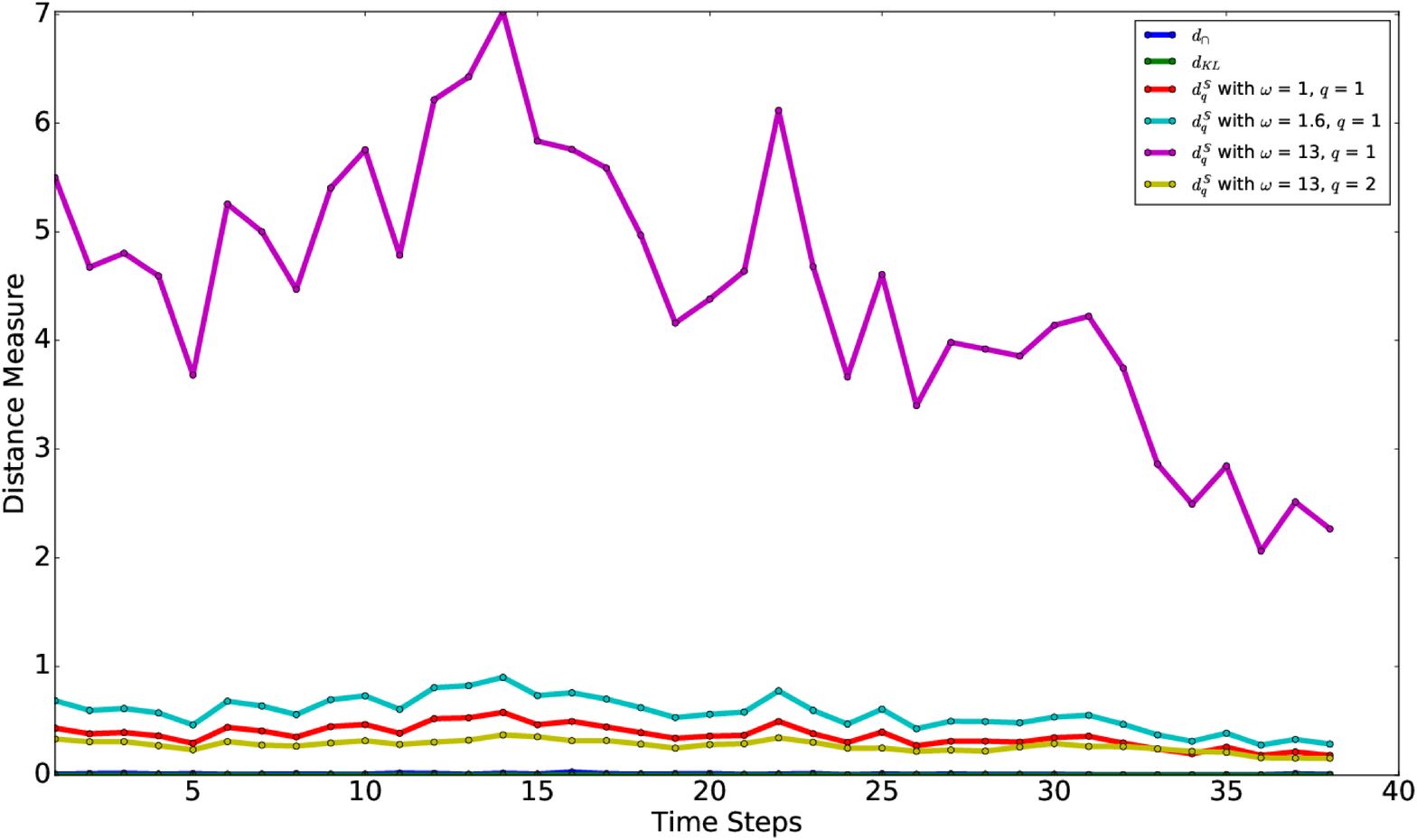} }}~~
    \hspace*{-.5cm}
    \subfloat[]{{\includegraphics[width=6.5cm]{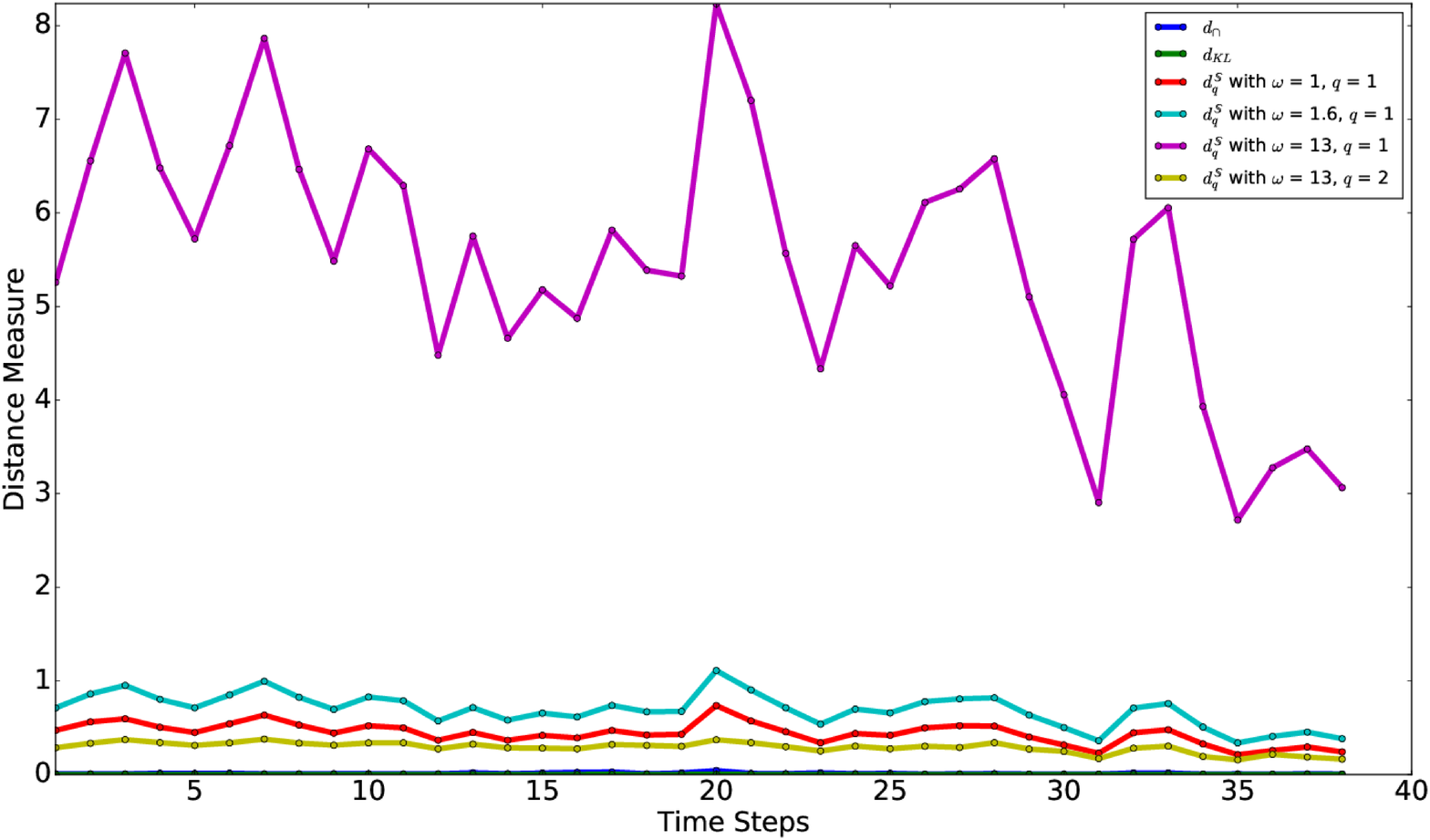} }}%
    \caption{Distance plot for scalar data for Pt-CO bond detection dataset. (a)~Plot for orbital density $69$ (HOMO$-1$). The highest peak is at site $16$. (b)~Plot for orbital density $70$ (HOMO). Significant peak is at site $21$.}%
    \label{fig:chemistry-scalar-plots}%
    \end{center}
\end{figure*}

We now describe an experiment to demonstrate the importance of studying tools for multifield data over single scalar field analysis tools. 
Consider the Pt-CO molecular dataset.
Using only orbital $69$ (HOMO-1) data the highest peak in the distance measure plot is obtained at site $16$ (Figure~\ref{fig:chemistry-scalar-plots}. Distance plots for orbital $70$ (HOMO) exhibit the highest peak at site $21$. 
On the other hand, using two fields together, i.e. orbital data $69$ and $70$, or orbital data $70$ and $71$, or orbital data $69$ and $71$, we observe the highest peak is always at site $21$. Some topological changes may not be captured using a bivariate data and we may need to consider more than two fields to detect the changes. 

\section{Conclusions and Future Work}
\label{sec:conclusion}
We propose the use of fiber-component distribution as a topological feature-descriptor for multifield data. We describe a novel method for extracting topological features from time-varying multifield data based on a distance measure defined between fiber-component distributions. This method is simple and a first step towards the development of a more accurate topological comparison measure between two Reeb spaces. We show effectiveness of our method by applying it on several datasets, both synthetic and real data. While the method captures important changes, it flags a few unimportant ones also. For example, in the plot for the Pt-Co data, we observe additional peaks. Such false positives are a key drawback of the current method. To overcome such issues in future we want to explore distance measures  between two Reeb Spaces. Overall, the proposed distance measures can be used to quickly identify interesting time-steps and intervals. The Reeb space could be studied in a subsequent step for detailed analysis. The distance measure can also be computed for sub-domains thereby allowing for finer grained analysis.\\

\noindent
{\textbf{Acknowledgments.}} 
We would like to thank all authors of the paper~\cite{2012-Duke-VisWeek} for sharing the nuclear scission data.
We also thank Prof. Brijesh Kumar Mishra,  IIIT-Bangalore for his expert help in understanding the Pt-CO interaction data. 
This work is partially supported by the Science and Engineering Research Board, India (SERB/CRG/2018/000702), IIIT-Bangalore, Department of Science and Technology, India (DST/SJF/ETA-02/2015-16), a Mindtree Chair research grant, and the Robert Bosch Centre for Cyber Physical Systems, Indian Institute of Science. 

\bibliographystyle{abbrv}
\bibliography{template}

\begin{thebibliography}{10}

\bibitem{1995-Gasteiger-H2-CO}
H.~A.~Gasteiger, N.~Markovi\.c, and P.~Ross.
\newblock H2 and co electrooxidation on well-characterized pt, ru, and pt-ru.
  2. rotating disk electrode studies of co/h2 mixtures at 62 .degree.c.
\newblock {\em The Journal of Physical Chemistry}, 99, 11 1995.

\bibitem{Bachthaler:2008:CS:1477066.1477444}
S.~Bachthaler and D.~Weiskopf.
\newblock Continuous scatterplots.
\newblock {\em IEEE Transactions on Visualization and Computer Graphics},
  14(6):1428--1435, Nov. 2008.

\bibitem{Bajaj:1997:CS:266989.267051}
C.~L. Bajaj, V.~Pascucci, and D.~R. Schikore.
\newblock The contour spectrum.
\newblock In {\em Proceedings of the 8th Conference on Visualization '97}, VIS
  '97, pages 167--ff., Los Alamitos, CA, USA, 1997. IEEE Computer Society
  Press.

\bibitem{bauer2014measuring}
U.~Bauer, X.~Ge, and Y.~Wang.
\newblock Measuring distance between reeb graphs.
\newblock In {\em Proceedings of the thirtieth annual symposium on
  Computational geometry}, page 464. ACM, 2014.

\bibitem{beketayev2014measuring}
K.~Beketayev, D.~Yeliussizov, D.~Morozov, G.~H. Weber, and B.~Hamann.
\newblock Measuring the distance between merge trees.
\newblock In {\em Topological Methods in Data Analysis and Visualization III},
  pages 151--165. Springer, 2014.

\bibitem{Biasotti2008DescribingSB}
S.~Biasotti, L.~D. Floriani, B.~Falcidieno, P.~Frosini, D.~Giorgi, C.~Landi,
  L.~Papaleo, and M.~Spagnuolo.
\newblock Describing shapes by geometrical-topological properties of real
  functions.
\newblock {\em ACM Comput. Surv.}, 40:12:1--12:87, 2008.

\bibitem{2007-Bremer}
P.-T. Bremer, E.~M. Bringa, M.~Duchaineau, D.~Laney, A.~Mascarenhas, and
  V.~Pascucci.
\newblock Topological {F}eature {E}xtraction and {T}racking.
\newblock {\em Journal of Physics: Conference Series}, 78:012007, 2007.

\bibitem{4015490}
H.~Carr, B.~Duffy, and B.~Denby.
\newblock On histograms and isosurface statistics.
\newblock {\em IEEE Transactions on Visualization and Computer Graphics},
  12(5):1259--1266, Sep. 2006.

\bibitem{JCN_paper}
H.~{Carr} and D.~{Duke}.
\newblock Joint contour nets.
\newblock {\em IEEE Transactions on Visualization and Computer Graphics},
  20(8):1100--1113, Aug 2014.

\bibitem{2015-Carr-Fiber}
H.~Carr, Z.~Geng, J.~Tierny, A.~Chattopadhyay, and A.~Knoll.
\newblock Fiber surfaces: Generalizing isosurfaces to bivariate data.
\newblock {\em Computer Graphics Forum}, 34(3):241--250, 2015.

\bibitem{2014-EuroVis-short}
A.~Chattopadhyay, H.~Carr, D.~Duke, and Z.~Geng.
\newblock {Extracting {J}acobi {S}tructures in {R}eeb {S}paces}.
\newblock In N.~Elmqvist, M.~Hlawitschka, and J.~Kennedy, editors, {\em EuroVis
  - Short Papers}, pages 1--4. The Eurographics Association, 2014.

\bibitem{2015-Chattopadhyay-CGTA-simplification}
A.~Chattopadhyay, H.~Carr, D.~Duke, Z.~Geng, and O.~Saeki.
\newblock Multivariate topology simplification.
\newblock {\em Computational Geometry: Theory and Application}, 58:1--24, 2016.

\bibitem{2012-Duke-VisWeek}
D.~Duke, H.~Carr, N.~Schunck, H.~A. Nam, and A.~Staszczak.
\newblock Visualizing {N}uclear {S}cission {T}hrough a {M}ultifield {E}xtension
  of {T}opological {A}nalysis.
\newblock {\em IEEE Transactions on Visualization and Computer Graphics},
  18(12):2033--2040, 2012.

\bibitem{2004-edels-jacobi}
H.~Edelsbrunner and J.~Harer.
\newblock Jacobi {S}ets of {M}ultiple {M}orse {F}unctions.
\newblock {\em In Foundations of Computational Matematics, Minneapolis, 2002},
  pages 37--57, 2004.
\newblock Cambridge Univ. Press, 2004.

\bibitem{2004-edels-localglobal}
H.~Edelsbrunner, J.~Harer, V.~Natarajan, and V.~Pascucci.
\newblock Local and {G}lobal {C}omparison of {C}ontinuous {F}unctions.
\newblock In {\em Proceedings of the conference on Visualization}, pages
  275--280, 2004.

\bibitem{2008-edels-reebspace}
H.~Edelsbrunner, J.~Harer, and A.~K. Patel.
\newblock Reeb {S}paces of {P}iecewise {L}inear {M}appings.
\newblock In {\em SoCG}, pages 242--250, 2008.

\bibitem{gao2010survey}
X.~Gao, B.~Xiao, D.~Tao, and X.~Li.
\newblock A survey of graph edit distance.
\newblock {\em Pattern Analysis and applications}, 13(1):113--129, 2010.

\bibitem{Dagstuhl-2014}
C.~Hansen, M.~Chen, C.~Johnson, A.~Kaufman, and H.~Hagen, editors.
\newblock {\em Scientific {V}isualization}.
\newblock Mathematics and Visualization. Springer-Verlag London, London, 2014.

\bibitem{Hardy-1952}
G.~H. Hardy, J.~E. Littlewood, and G.~P\'olya.
\newblock {\em Inequalities}.
\newblock Cambridge: Cambridge University Press. ISBN 0-521-35880-9, second
  edition, 1952.

\bibitem{hilaga2001topology}
M.~Hilaga, Y.~Shinagawa, T.~Kohmura, and T.~L. Kunii.
\newblock Topology matching for fully automatic similarity estimation of 3d
  shapes.
\newblock In {\em Proceedings of the 28th annual conference on Computer
  graphics and interactive techniques}, pages 203--212. ACM, 2001.

\bibitem{2013-Huettenberger-pareto}
L.~Huettenberger, C.~Heine, H.~Carr, G.~Scheuermann, and C.~Garth.
\newblock Towards {M}ultifield {S}calar {T}opology {B}ased on {P}areto
  {O}ptimality.
\newblock {\em Computer Graphics Forum}, 32(3.3):341--350, 2013.

\bibitem{Ji_featuretracking}
G.~Ji and H.~wei Shen.
\newblock Feature tracking using earth movers distance and global
  optimization,” pacific graphics 2006.

\bibitem{2010-JACS-Kendric}
I.~Kendrick, D.~Kumari, A.~Yakaboski, N.~Dimakis, and E.~Smotkin.
\newblock Elucidating the ionomer-electrified metal interface.
\newblock {\em Journal of the American Chemical Society}, 132, 11 2010.

\bibitem{kitware2003}
Kitware, Inc.
\newblock {\em The Visualization Toolkit User's Guide}, January 2003.

\bibitem{2003-Kresse-Pt-CO}
G.~Kresse, A.~Gil, and P.~Sautet.
\newblock Significance of single-electron energies for the description of co on
  pt(111).
\newblock {\em Phys. Rev. B}, 68, 08 2003.

\bibitem{Lee_TAcbased}
T.-Y. Lee and H.-W. Shen.
\newblock Visualizing time-varying features with tac-based distance fields.
\newblock In {\em 2009 IEEE Pacific Visualization Symposium}, 2009.

\bibitem{Lehmann-2010}
D.~J. {Lehmann} and H.~{Theisel}.
\newblock Discontinuities in continuous scatter plots.
\newblock {\em IEEE Transactions on Visualization and Computer Graphics},
  16(6):1291--1300, Nov 2010.

\bibitem{morozov2013interleaving}
D.~Morozov, K.~Beketayev, and G.~Weber.
\newblock Interleaving distance between merge trees.
\newblock {\em Discrete and Computational Geometry}, 49(22-45):52, 2013.

\bibitem{narayanan2015distance}
V.~Narayanan, D.~M. Thomas, and V.~Natarajan.
\newblock Distance between extremum graphs.
\newblock In {\em 2015 IEEE Pacific Visualization Symposium (PacificVis)},
  pages 263--270. IEEE, 2015.

\bibitem{rubner2000earth}
Y.~Rubner, C.~Tomasi, and L.~J. Guibas.
\newblock The earth mover's distance as a metric for image retrieval.
\newblock {\em International journal of computer vision}, 40(2):99--121, 2000.

\bibitem{2004-Saeki}
O.~Saeki.
\newblock {\em Topology of {S}ingular {F}ibers of {D}ifferentiable {M}aps}.
\newblock Springer, 2004.

\bibitem{Saeki2014}
O.~Saeki, S.~Takahashi, D.~Sakurai, H.-Y. Wu, K.~Kikuchi, H.~Carr, D.~Duke, and
  T.~Yamamoto.
\newblock {\em Visualizing {M}ultivariate {D}ata {U}sing {S}ingularity
  {T}heory}, volume~1 of {\em Mathematics for Industry}, chapter The Impact of
  Applications on Mathematics, pages 51--65.
\newblock Springer Japan, 2014.

\bibitem{saikia2014extended}
H.~Saikia, H.-P. Seidel, and T.~Weinkauf.
\newblock Extended branch decomposition graphs: Structural comparison of scalar
  data.
\newblock {\em Computer Graphics Forum}, 33(3):41--50, 2014.

\bibitem{saikia2015fast}
H.~Saikia, H.-P. Seidel, and T.~Weinkauf.
\newblock Fast similarity search in scalar fields using merging histograms.
\newblock In {\em Topological Methods in Data Analysis and Visualization},
  pages 121--134. Springer, 2015.

\bibitem{scott1988note}
D.~W. Scott.
\newblock A note on choice of bivariate histogram bin shape.
\newblock {\em Journal of Official Statistics}, 4(1):47--51, 1988.

\bibitem{book-Surface-Chem}
G.~A. Somorjai and Y.~Li.
\newblock {\em Introduction to Surface Chemistry and Catalysis}.
\newblock John Wiley \& Sons, 2010.

\bibitem{sridharamurthyedit}
R.~{Sridharamurthy}, T.~{Bin Masood}, A.~{Kamakshidasan}, and V.~{Natarajan}.
\newblock Edit distance between merge trees.
\newblock {\em IEEE Transactions on Visualization and Computer Graphics}, pages
  1--1, 2018.

\bibitem{thomas2014multiscale}
D.~M. Thomas and V.~Natarajan.
\newblock Multiscale symmetry detection in scalar fields by clustering
  contours.
\newblock {\em IEEE Trans. Visualization and Computer Graphics},
  20(12):2427--2436, 2014.

\end{thebibliography}

\end{document}